\documentclass[aps,prx,floatfix,twocolumn,showpacs,superscriptaddress]{revtex4-1}
\usepackage{graphicx}
\usepackage{amssymb}
\usepackage{amsmath}
\usepackage{soul}
\usepackage[usenames]{color}
\usepackage{ulem}

\ifx\pdftexversion\undefined
\usepackage[dvips]{hyperref}
\else
\usepackage{hyperref}
\fi
\hypersetup{
  colorlinks = true, linkcolor = blue
}

\begin{document}
\title{Three-Dimensional Laser Cooling at the Doppler limit}

\author{R. Chang} 
\author{A. L. Hoendervanger}
\author{Q. Bouton} 
\affiliation{Laboratoire Charles Fabry, Institut d'Optique, CNRS, Univ. Paris Sud, 2 Avenue Augustin Fresnel 91127 PALAISEAU cedex, France}
\author{Y. Fang}
\affiliation{Laboratoire Charles Fabry, Institut d'Optique, CNRS, Univ. Paris Sud, 2 Avenue Augustin Fresnel 91127 PALAISEAU cedex, France}
\affiliation{Quantum Institute for Light and Atoms, Department of Physics, State Key Laboratory of Precision Spectroscopy, East China Normal University, Shanghai, 200241, China}
\author{T. Klafka}
\affiliation{Laboratoire Charles Fabry, Institut d'Optique, CNRS, Univ. Paris Sud, 2 Avenue Augustin Fresnel 91127 PALAISEAU cedex, France}
\author{K. Audo}
\author{A. Aspect}
\author{C. I. Westbrook}
\author{D. Cl\'ement}
\affiliation{Laboratoire Charles Fabry, Institut d'Optique, CNRS, Univ. Paris Sud, 2 Avenue Augustin Fresnel 91127 PALAISEAU cedex, France}

\begin{abstract}
Many predictions of Doppler cooling theory of two-level atoms have never been verified in a three-dimensional geometry, including the celebrated minimum achievable temperature $\hbar \Gamma/2 k_B$, where $\Gamma$ is the transition linewidth. Here, we show that, despite their degenerate level structure, we can use Helium-4 atoms to achieve a situation in which these predictions can be verified. We make measurements of atomic temperatures, magneto-optical trap sizes, and the sensitivity of optical molasses to a power imbalance in the laser beams, finding excellent agreement with the Doppler theory.  We show that the special properties of Helium, particularly its small mass and narrow transition linewidth, prevent effective sub-Doppler cooling with red-detuned optical molasses.
\end{abstract}

\maketitle

\section{Introduction}

The seminal proposals for Doppler cooling in 1975  \cite{Hansch1975,Wineland1975} prompted the realization of the first optical molasses in the 1980's \cite{Chu1985}, a major landmark in the field of laser cooling and trapping of atoms \cite{cohen1998nobel, chu1998manipulation, Phillips1998ki}. The physical concepts behind the Doppler cooling mechanism are both simple and elegant, and predict the achievement of very low temperatures. They remain to this day the starting point of most courses on laser cooling and degenerate gases \cite{Cohen1992}. It is thus ironic that not only, to our knowledge, quantitative predictions of this simple model have not been experimentally validated, but moreover, experimental studies have found results quantitatively and even qualitatively different from the predictions of the model \cite{Lett1988,Lett1989,Weiss1989,Salomon1990}. 

Today, it is well known that for most atoms with a degenerate ground state manifold, the complex multi-level structure gives rise to new mechanisms which come to dominate the cooling process, yielding ultimate temperatures far below those predicted by Doppler theory \cite{Dalibard1989,Ungar1989,Cohen1992}. Accounting for these sub-Doppler mechanisms allows one to understand the qualitative and quantitative experimental observations.  In contrast, three-dimensional laser cooling of an atomic sample in agreement with the celebrated Doppler model is still lacking \cite{Note1DCooling}.  Recent work with alkaline-earth and rare-earth atoms, which exhibit a non-degenerate ground state, have provided natural candidates to study cooling in a purely Doppler regime.  Yet, to our knowledge, all experiments  with these atoms have found temperatures above the expected Doppler limit in magneto-optical traps (MOTs) \cite{Zinner2000,Curtis2001,Grunert2001, Dammalapati2009,Binnewies2001,Kisters1994,Sengstock1994,Riedman2012, Katori1999,Xu2002, Kuwamoto1999,Loftus2000,Dorscher2013}. This discrepancy has been attributed to the presence of additional heating mechanism \cite{Chaneliere2005,Choi2008}.  

\begin{figure}[h!]
\begin{center}
\includegraphics[width= 0.8\columnwidth]{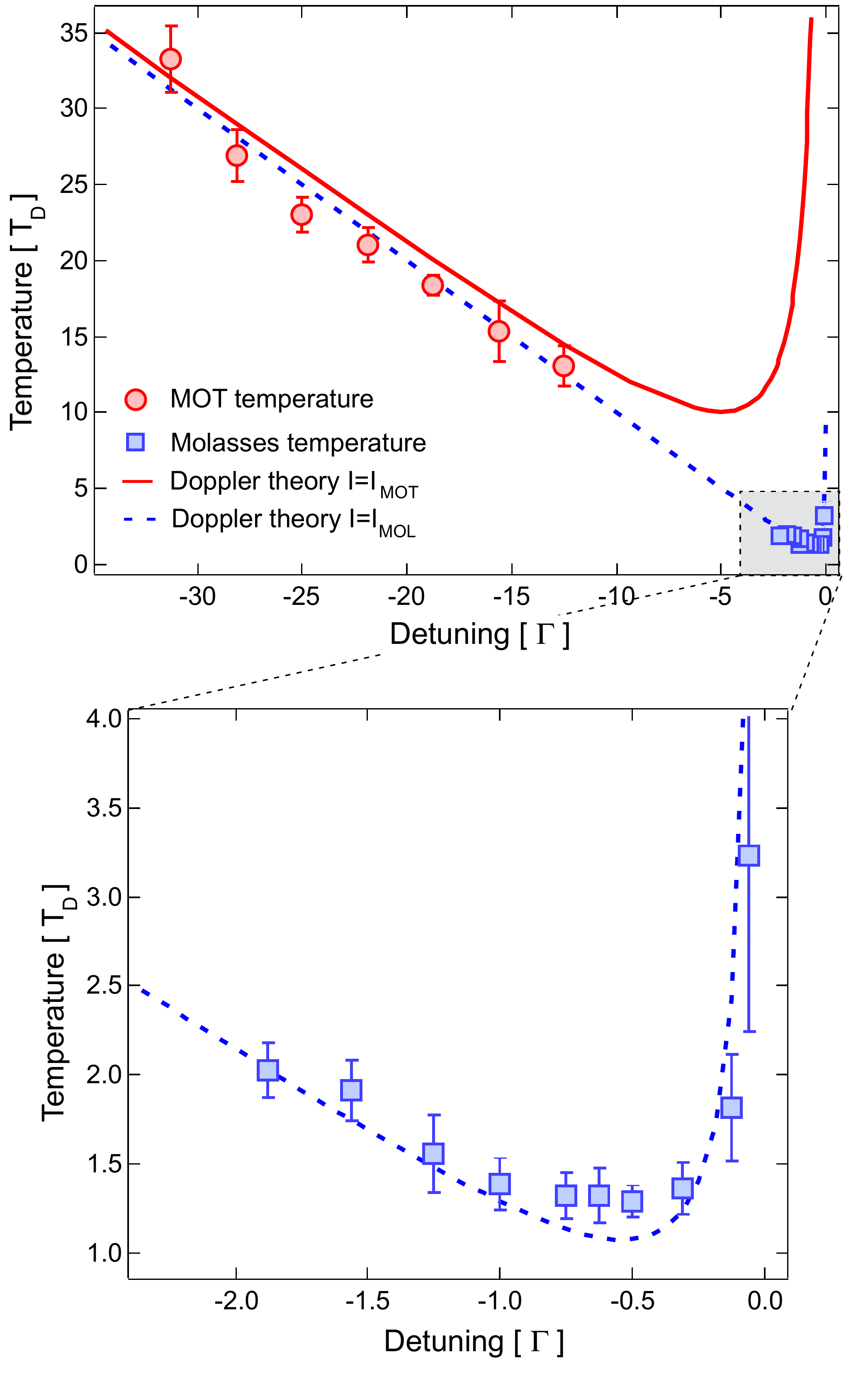}%
\caption{Temperature in magneto-optical trap and optical molasses as a function of the detuning of the laser cooling beams. Temperatures are extracted from monitoring time-of-flight expansion of the gas. Error bars (one standard deviation) reflect the error from fitting. The total intensity (sum of the six beams) used in the experiment are respectively $I_{{\rm MOT}}=100\ I_{{\rm sat}}$ and $I_{{\rm MOL}}=I_{{\rm sat}}/10$. The lines are the results of  Doppler theory of Eq.~\ref{Eq:Tdoppler}, for intensities $I_{{\rm MOT}}$ (solid line) and $I_{{\rm MOL}}$ (dashed line).}
\label{Fig:DopplerMOT-Mol}%
\end{center}
\end{figure}

Here we report the three-dimensional laser cooling of metastable Helium-4 gases in the Doppler regime.  We monitor the temperature of the gas as a function of the laser detuning $\delta$, as shown in Fig.~\ref{Fig:DopplerMOT-Mol}, finding excellent agreement with Doppler theory.  The temperature reaches a minimum for a detuning $\delta=- \Gamma / 2$ (where $\Gamma$ is the transition linewidth), while increasing with $|\delta|$, in contrast with sub-Doppler molasses. In addition, the drift velocities of optical molasses with unbalanced power between counter-propagating beams are far larger than those expected of sub-Doppler molasses \cite{Lett1989}.  We use the $2^3S_{1} \rightarrow 2^3P_{2}$ transition of Helium-4, which allows in principle for sub-Doppler cooling, and yet our results show no evidence for such effects.  We present a physical argument showing that the special properties of metastable Helium ($^4$He*) strongly inhibit sub-Doppler cooling in the experimental configuration we probe.

\section{Theoretical considerations} \label{sec:theory}

\subsection{Laser cooling mechanisms}

Laser cooling of an atomic gas relies upon the exchange of momentum between the atoms and the near-resonant light field, resulting in a mechanical force $F$ on the atoms.  For small velocities, the equilibrium temperature $T$ of these cooling schemes is given by the ratio of a momentum-space diffusion constant $\mathcal{D}$ (given by the fluctuations of the force) to the velocity-damping coefficient $\alpha$.  In the following we will first recall the main theoretical results for laser Doppler cooling of a two-level atom, then proceeding to discuss a multi-level atom where sub-Doppler mechanisms may appear.

The mechanical interaction of a near-resonant light beam (frequency $\omega/2\pi$) with a two-level atom (atomic resonance frequency $\omega_{0}/2\pi$) is dominated by the radiation pressure effect and its associated force \cite{Lett1989}.  This light-matter interaction is characterised by the laser detuning from resonance $\delta=\omega-\omega_{0}$, the laser intensity $I$, the natural linewidth of the transition $\Gamma$, and the saturation intensity of the transition $I_{sat}$.  A convenient physical quantity is the generalized saturation parameter
\begin{equation}
s=\frac{s_0}{1+4\delta^2/\Gamma^2},
\end{equation}
which characterizes the excited state population in a 2-level atom.  Here $s_0=I/I_{sat}=2\Omega^2/\Gamma^2$ is the on-resonance saturation parameter, and $\Omega$ is the Rabi frequency.

\begin{figure*}
\begin{center}
\includegraphics[width=1.7 \columnwidth]{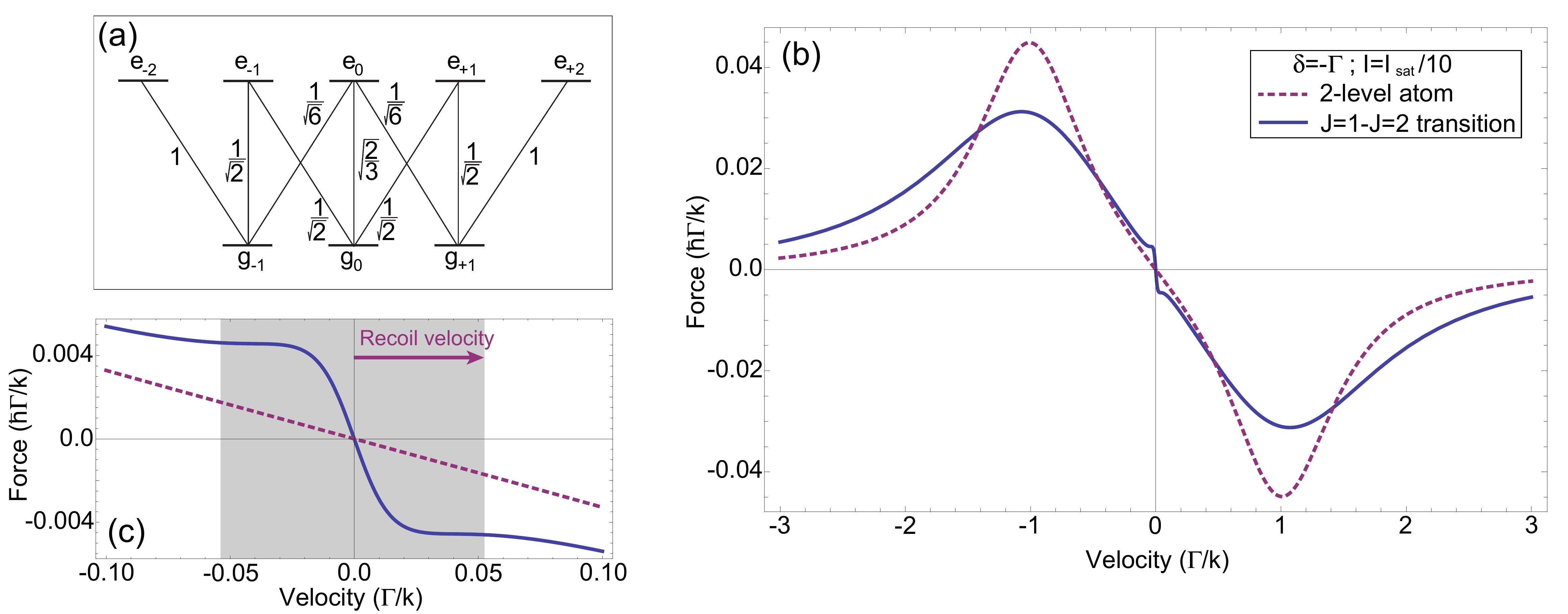}%
\caption{{\bf (a)} Atomic structure of a multi-level atom on a $J=1 \rightarrow J'=2$ transition. {\bf (b)} Semi-classical force for a moving Helium atom in a $\sigma^{+}-\sigma^{-}$ standing wave in the case of a two-level atom (dashed line, calculated as in \cite{Lett1989}) and in that of a multi-level atom on a $J=1 \rightarrow J'=2$ transition (solid line, calculated as in \cite{Chang1999}). Detuning and intensity correspond to typical values used in the experiment for optical molasses. {\bf (c)} Expanded view of the low velocity region of the semi-classical forces, high-lighting the narrow range of the sub-Doppler feature.  In both calculations, $s=0.02.$}%
\label{Fig:CompForces}%
\end{center}
\end{figure*}

At low saturation of the atomic transition $s \ll 1$  and small velocities $ |kv|\ll |\delta|$, the average force acting on a two-level atom moving at velocity $v$ takes the form $F_{\rm 2 level}=-\alpha_{\rm 2 level}v$ with
\begin{equation}
\alpha_{\rm 2 level} = - 4 \hbar k^2 \frac{2 \delta/\Gamma \ 2 \Omega^2/\Gamma^2 }{(1+ 2 \Omega^2/\Gamma^2+(2 \delta/\Gamma)^2)^2}.
\label{Eq:alpha}
\end{equation}
Here $k$ is the photon wavevector, and $\hbar$ the reduced Planck's constant. The force $F_{\rm 2 level}$ is plotted in Figure~\ref{Fig:CompForces} (dashed line) for the typical experimental parameters of optical molasses.

Calculating the diffusion constant $\mathcal{D}_{\rm 2level}$ for the two-level atom, one can obtain the equilibrium temperature $k_{B} T_{\rm 2level}=\mathcal{D}_{\rm 2level}/\alpha_{\rm 2level}$, where $k_{B}$ is Boltzmann's constant. Following the generalization to cooling in 3D \cite{Lett1989}, the temperature of a 3D molasses formed by three orthogonal pairs of counter-propagating laser beams is,
\begin{equation}
k_{B}  T_{\rm 2 level} =\frac{\hbar \Gamma}{2} \frac{1 + I_{{\rm tot}}/I_{{\rm sat}} +(2 \delta/\Gamma)^2}{4 |\delta|/\Gamma},
\label{Eq:Tdoppler}
\end{equation}
where the $I_{{\rm tot}}=6 I$ is the total intensity resulting from all six laser beams.  In the Doppler regime, the minimum possible temperature is achieved for a detuning $\delta=-\Gamma/2$ and vanishing intensity.  This minimum is commonly referred to as the Doppler limit, $T_{D}=\hbar \Gamma/2k_B$ .

We now turn the discussion to an atom with several ground-state levels where sub-Doppler mechanisms of cooling may be involved \cite{Dalibard1989, Lett1989, Weiss1989}. For this discussion we restrict ourselves to the atomic structure of the $^4$He* atom on the $2^3S_{1} \rightarrow 2^3P_{2}$ transition (see Fig.~\ref{Fig:CompForces}{\color{blue}a}), and the laser configuration used in the experiment: counter-propagating laser beams with opposite circular polarization (configuration $\sigma^+-\sigma^-$) along three orthogonal axes.

Consider first a single axis of the system. Following the derivation from \cite{Chang1999} an explicit formulation of the semi-classical force $F$ on a moving atom can be obtained. This approach takes into account atomic coherence effects up to second-order in absorption-emission processes. In Fig.~\ref{Fig:CompForces} we plot this force $F$ along with that exerted on a two-level atom $F_{\rm 2level}$. In the low-velocity region (see Fig.~\ref{Fig:CompForces}{\color{blue}c}) the cooling force on the multi-level atom exhibits a sharp feature with an associated damping coefficient $\alpha$ significantly larger than that of a two-level atom.  This feature arises from the presence of atomic coherences and two-photon processes, and is the signature of sub-Doppler cooling mechanisms.  Since the momentum-space diffusion coefficient hardly changes, the new damping coefficient would lead one to expect sub-Doppler cooling.  

In this configuration, the additional cooling results from a motion-induced population difference between the ground-state levels, and is referred to as  $\sigma^+-\sigma^-$ polarization gradient cooling.  In a 3D configuration, it is well known that Sisyphus cooling can also occur due to spatially dependent light shifts of the atomic ground-state levels.  The electric field from laser beams along orthogonal axes interfere, resulting in a modulation of the intensity, which in turn may lead to Sisyphus cooling as in a one-dimensional lin $\perp$ lin configuration.  Within the parameter range of interest ($|\delta|\sim\Gamma$), both sub-Doppler effects result in similar velocity capture ranges and equilibrium temperatures \cite{Dalibard1989, Lett1989, Weiss1989}.

The comparison between a 2-level and a multi-level atom presented in Fig.~\ref{Fig:CompForces} suggests two important conclusions.  On the one hand the two-photon structure occurs only in a small range close to zero velocity (see Fig.~\ref{Fig:CompForces}{\color{blue}c}). In particular, for typical experimental parameters with Helium-4 atoms this velocity range is smaller than the recoil velocity $v_R=\hbar k/m$, where $m$ is the particle's mass. This implies that sub-Doppler cooling is highly ineffective in a red molasses of $^4$He* since the velocity capture range for such processes is much smaller than the lowest velocity achievable through Doppler cooling. On the other hand, on a larger velocity range corresponding to Doppler velocities the friction coefficient for the multi-level atom is close to that of an effective two-level atom.  In the following section, we will study the sub-Doppler capture velocity range in greater detail, showing that indeed $^4$He* occupies a special place in the study of laser cooling mechanisms.

\subsection{Capture range of sub-Doppler cooling}
The velocity capture range for Doppler cooling is $v_{c}^D \simeq |\delta|/k$, directly reflecting the largest Doppler effect that can be compensated with a detuning $\delta$ from resonance. Similarly, the capture range for sub-Doppler cooling reflects the new physical mechanisms involved.  For Sisyphus cooling (lin $\perp$ lin) to be effective, a ground-state atom should have a high probability of being optically pumped into a different ground-state atom on a wavelength distance \cite{Cohen1992}.  In the semi-classical regime, this yields a condition on the atomic velocity $v \ll \Gamma'/k$, where $\Gamma'$ is the optical pumping rate from one ground-state level to another ground-state level.  The capture velocity is thus estimated to be $v_{c}^{\perp } \simeq \Gamma'/k$.  In the case of $\sigma^+-\sigma^-$ polarization gradient cooling, the capture velocity is given by the pumping rate at small detunings $\delta \sim - \Gamma$, $v_{c}^{\sigma} \simeq \Gamma'/k$ \cite{Chang1999}, while it is set by the light shift $\delta'$ of the ground-state levels at large detuning $|\delta| \gg \Gamma$, and is defined in the same way to be $v_{c}^{\sigma} \simeq |\delta'|/k$, \cite{Cohen1992}.  The optical pumping rate and the light shift can be expressed conveniently as \cite{Cohen1992}

\begin{gather}
\Gamma' = \frac{\Gamma}{2} \frac{ 2\Omega^2/\Gamma^2}{1+ 2\Omega^2/\Gamma^2+(2\delta/\Gamma)^2} =  \frac{\Gamma}{2} \frac{s}{1+s}  \nonumber \\
\delta' = \frac{\delta}{2} \frac{  2\Omega^2/\Gamma^2}{1+ 2\Omega^2/\Gamma^2+(2\delta/\Gamma)^2} =  \frac{\delta}{2} \frac{s}{1+s}  \nonumber \\
\label{Eq:PumpingRate}
\end{gather} 

A large capture velocity range for sub-Doppler cooling requires a large $\Gamma'$ and/or $\delta'$. These two quantities are decreasing functions of the detuning $\delta$ when $|\delta |\gg \Gamma$ and $\delta'$ is maximum for $|\delta|=\Gamma/2$. The Doppler temperature (see Eq.~\ref{Eq:Tdoppler}) is minimum in the same range of detuning $\delta$. In the following, we will therefore concentrate on laser detuning $|\delta|$ of the order of $\Gamma$, the most favorable situation to observe signatures of sub-Doppler cooling (although not the lowest temperature). The capture velocities for both sub-Doppler mechanisms are then of the same order,
\begin{gather}
v_{c}^{\perp} \simeq v_{c}^{\sigma} = v_{c} =   \frac{\Gamma}{2k}  \frac{s}{1+s}.
\label{Eq:VCapture}
\end{gather}

For the parameters represented in Fig.~\ref{Fig:CompForces} ($s=0.02$) and using Eq.~\ref{Eq:VCapture}, we find $v_{c}\simeq 0.01 \Gamma/k$ in agreement with the numerical calculation of the force shown in Fig.~\ref{Fig:CompForces}. Identifying a root-mean-square (RMS) velocity $v_{D}$ of a 3D gas at the Doppler temperature $T_{D}$, $\frac{1}{2} m v_{D}^2=\frac{3}{2} k_{B} T_{D}$,  we find
\begin{gather}
 \frac{v_{c}}{v_{D}}=\sqrt{\frac{1 }{6 } \frac{ T_{D}}{ T_{R}} } \frac{s}{1+s},
\label{Eq:CaptureFraction}
\end{gather}
where $T_{R}$ is the temperature associated with the recoil energy $k_BT_{R}=\hbar^2k^2/2m$.  The capture velocity for sub-Doppler cooling is proportional to $\Gamma$ (via $v_{D}$) and the square-root of the ratio $T_{D}/T_{R}$.  Both these quantities are small in the case of $^4$He* as compared to other species. In the Table \ref{TableCompTc} we present a comparison of the ratio $T_{D}/T_{R}$ for several atomic species commonly used in laser-cooling experiments. 

\begin{table}
\setlength{\tabcolsep}{2.5 mm}
\renewcommand{\arraystretch}{1.1}
\begin{tabular}{|c|c|c|}
  \hline
  \ Species \ &  \ $T_{D}/T_{R}$  &  \ $v^{{\rm up}}_{c}/v_{D}$ \ \\
  \hline
  $^4$He* & 18.8  & 0.59   \\
  $^7$Li & 39.4 & 0.85 \\
  $^{23}$Na & 196 & 1.90 \\
  $^{40}$K & 348 & 2.54\\
  $^{87}$Rb & 808   & 3.87 \\
  $^{133}$Cs & 1413  & 5.12\\
  \hline
\end{tabular}
\caption{Ratio $T_{D}/T_{R}$ and estimated maximum capture velocity $v^{{\rm up}}_{c}/v_{D}$ ($s_{max}=1/2$) for several commonly laser-cooled atoms.}\label{TableCompTc}
\end{table}

Eq.~\ref{Eq:CaptureFraction} indicates that the capture velocity  $v_{c}/v_{D}$ increases with saturation parameter $s$, approaching the limit $\sqrt{T_{D}/6 T_{R} }$.  However, $s$ can not be increased arbitrarily.  For red-detuned optical molasses, the friction coefficient $\alpha$ is known to change sign at large saturation parameter \cite{Minogin1979}. This change of sign occurs when the contribution of multi-photon processes to the force becomes large enough to modify the excited state decay rate, and typically occurs at $s \sim 1$.  Supposing a maximum saturation parameter $s_{max}=1/2$, we estimate an upper limit on the capture velocity $v^{up}_{c}/v_{D}$. The values for $v^{up}_{c}/v_{D}$ are given in Table~\ref{TableCompTc}.  It is clear that Helium, and to a lesser extent Lithium, is special with respect to other alkali atoms since the capture range for sub-Doppler cooling is smaller than the range of velocities achievable with Doppler cooling. This statement applies both to the $\sigma^+-\sigma^-$ and lin $\perp$ lin configurations.  Thus we conclude that sub-Doppler cooling is not expected to play a role on the $2^3S_{1}\rightarrow 2^3P_{2}$ transition of Helium-4.  Indeed, in the experiment we do not observe any signatures of sub-Doppler cooling.  The semi-classical arguments we have presented above highlight the special place that Helium occupies among laser cooled species.  A precise and quantitative condition for the appearance of efficient sub-Doppler cooling would require a fully quantum, three-dimensional computation and is beyond the scope of this paper. 

\subsection{Equilibrium temperature for a multi-level atom in the Doppler regime}

From the above arguments, sub-Doppler effects on metastable Helium on the $2^3S_{1} \rightarrow 2^3P_{2}$ transition are expected to be negligible.  We will therefore compare the temperature measurements in 3D gases to the predictions of Doppler theory \cite{Lett1989}.  To account for the multi-level atomic structure in the 3D Doppler theory we take a weighted sum over all possible one-photon transitions, where the weights are given by the square of the Clebsch-Gordan coefficients  (see Fig.~\ref{Fig:CompForces}{\color{blue}a}).  This leads to a rescaling of the saturation intensity $I_{0}=9/5 \ I_{sat}$, and an equilbrium temperature
\begin{equation}
k_{B} T =\frac{\hbar \Gamma}{2} \frac{1 + I_{tot}/I_{0} +(2 \delta/\Gamma)^2}{4 |\delta|/\Gamma}.
\label{Eq:Tdoppler}
\end{equation}
where $I_{tot}$ is the total intensity in the 6 beams.  A similar approach has been used in \cite{Townsend1995}.

\section{Description of the experimental apparatus}\label{sec:apparatus}

\subsection{$^4$He* Magneto-Optical traps}

Our measurements are performed with an apparatus which cools and traps metastable Helium atoms in a MOT. $^4$He* atoms are produced in a hot plasma (dc-discharge) and slowed down to trappable velocities on the order of several tens of meters per second with a $2.5~$m long Zeeman slower. The slowed atoms enter the science chamber where three orthogonal pairs of counter-propagating laser beams are shone onto the atoms in the presence of a quadrupole magnetic field.  The cooling light, which addresses the $2^3S_{1}\rightarrow 2^3P_{2}$ transition, is derived from a Koheras AdjustiK Y10 fiber laser from NKT Photonics with a manufacturer stated linewidth $<10$ kHz.  During the MOT phase, the typical intensity per beam is $\sim 20 I_{sat}$ at a detuning $\delta_{MOT}=-50~$MHz $\simeq -31 \Gamma$ from the atomic transition, where the transition linewidth is $\Gamma=2 \pi \times 1.6$MHz. The magnetic field gradient along the coil axis is $B'_x=24$ G/cm. Under these conditions, $8 \times 10^8$ atoms at a temperature of $1.5(1)$ mK are loaded within 2 seconds.

Detection of the gas is performed using a thermoelectrically cooled InGaAs camera (XEVA type from Xenics company -- $256 \times 320$ pixels with a pixel size of $30 \times 30~\mu$m). This technology is suited to image metastable Helium atoms with a quantum efficiency of $\sim 80~$\% at 1083 nm. The camera collects the fluorescence of the atoms from the probe beams. The latter are made of the six beams we use to make a MOT which are tuned onto resonance of the atomic transition during the imaging pulse. The duration of the imaging pulse is 100 $\mu$s and the total intensity is about 175 $I_{sat}$, where $I_{sat}= \pi h c \Gamma /3 \lambda^3 \simeq0.165$ mW/cm$^{2}$ is the saturation intensity of the $J=1\rightarrow J^\prime=2$ cycling transition. Sizes and temperature of the $^4$He* clouds are extracted by monitoring the time-of-flight expansion of the initially trapped gases and fitting the imaged 2D density profiles with a Gaussian function. 

\begin{figure*}[ht!]
\begin{center}
\includegraphics[width=2\columnwidth]{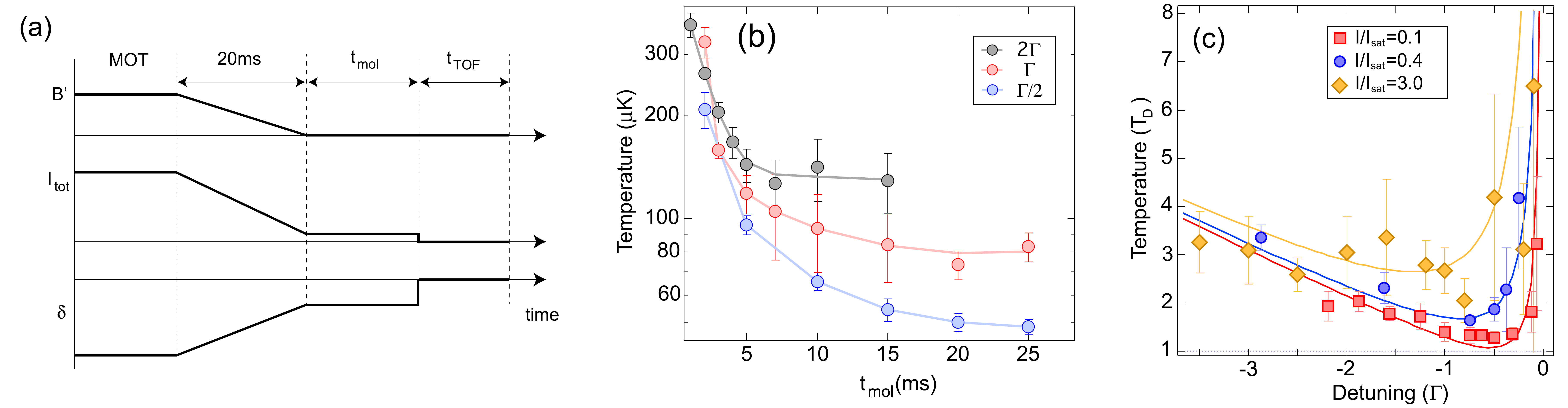}%
\caption{{\bf (a)} Sketch of the experimental cycle to probe optical molasses. After the MOT phase, the magnetic field gradient $B'$, the beam intensity $I_{tot}$ and the detuning $\delta$ are ramped over 20 ms.  A molasses phase at constant parameters lasts $t_{mol}$. After switching off the molasses beams, fluorescence pictures are taken after a time-of-flight $t_{TOF}$. {\bf (b)} Temperature of optical molasses as a function of the duration $t_{mol}$ of the second stage of molasses. Solid lines are guide to the eye. {\bf (c)} Temperature of optical molasses as a function of the laser detuning.  Comparison with Doppler theory for laser cooling multi-level atoms of Eq.~\ref{Eq:Tdoppler} (solid lines). The different sets correspond to different intensity in the cooling beams.}%
\label{Fig:EqTime}%
\end{center}
\end{figure*}

\subsection{Optical molasses}

After the MOT phase we implement an optical molasses on the $2^3S_{1} \rightarrow 2^3P_{2}$ transition, as we shall now describe. At the end of the MOT phase we ramp the magnetic field gradient to zero and ramp both the detuning and intensity of the laser beams from the MOT values to those of the molasses within 20 ms (see Fig.~\ref{Fig:EqTime}{\color{blue}a}). This ramp of the parameters allows us to capture and cool half of the atoms ($N=4 \times 10^8$) in the molasses. The polarization of the light beams during the molasses stage is identical to that of the MOT. We then wait for a variable time $t_{mol}$ at fixed final intensity and detuning of the laser beams to reach thermal equilibrium. We monitor the time-of-flight expansion of the optical molasses cloud to extract its temperature.

\section{Results}\label{sec:results}

\subsection{Temperature}

In Fig.~\ref{Fig:DopplerMOT-Mol} we present the results of the temperature measurements performed on the MOT and on the optical molasses by recording the TOF expansion of the atomic clouds. Temperature is plotted as a function of the laser detuning $\delta$. We observe a minimum temperature at a detuning $\delta=-\Gamma/2$ as predicted by Doppler cooling theory. The minimum measured temperature $T=1.3(1)$ $T_{D}$ is close to the expected Doppler limit, which occurs for vanishingly small light intensity.  In addition we compare our temperature measurements to the Doppler prediction of Eq.~\ref{Eq:Tdoppler} and we find excellent agreement over the entire range of detunings studied in the experiment.  For the high-temperature measurements, the confining potential provided by the MOT is necessary to reach thermal equilibrium on the timescale of the experiment.  On the other hand, the low-temperature measurements are performed on an optical molasses (see Figs.~\ref{Fig:DopplerMOT-Mol} and \ref{Fig:EqTime}{\color{blue}c}), since otherwise density-dependent Penning collisions would severely reduce the number of trapped atoms \cite{Shlyapnikov1994}.

\subsection{Equilibration time for 3D optical molasses}
In Fig.~\ref{Fig:EqTime}{\color{blue}b} we plot the time evolution of the atom cloud temperature during the molasses phase. On a short timescale of $t_{mol} \sim$ 1 ms, we observe a rapid decrease of the temperature.  However, reaching the equilibrium temperature requires durations even longer.  Close to the Doppler limit $T_{D}$ ($\delta=-\Gamma/2$ and $I_{tot}/I_{sat}=1/10$), the molasses temperature reaches equilibrium on a timescale $t_{mol} \simeq 10$ ms.  We have observed that the equilibration time can vary drastically in the presence of uncompensated bias magnetic fields and power imbalances between counter-propagating beams. As we will see in Section~\ref{ssec:driftVelocity}, the equilibirum molasses state is also very sensitive to such technical issues. 

In Doppler theory, the timescale to reach the equilibrium temperature is directly related to the velocity damping coefficient, predicting an expected cooling time of $\tau_{cool}=m/2\alpha$ \cite{Lett1989}. For the low value of the saturation parameter $s\simeq 0.05$ and detuning $|\delta|\sim\Gamma$ used in the data presented in Fig.~\ref{Fig:EqTime}, the expected Doppler cooling time is $t_{cool}\simeq 0.5$ ms.  Although this timescale is similar to that of the observed initial rapid decrease in temperature, for the above stated reasons we can not use this measurements to estimate $\alpha$.

\subsection{MOT sizes}

The Penning collision rate in non-polarized $^4$He* is relatively high \cite{Shlyapnikov1994}, limiting the atomic density of laser-cooled Helium clouds to $\sim 10^{9}$ cm$^{-3}$.  As a result, photon re-scattering effects, which can result in heating of the atom cloud in other species, are typically negligible for Helium \cite{NoteMultipleScatt}. As a consequence, the equilibrium of metastable Helium MOTs is reached in the absence of both multiple scattering and sub-Doppler cooling. In this simple regime, the equilibrium temperature  is given by Eq.~\ref{Eq:Tdoppler} and the MOT sizes derive from the equipartion theorem at the Doppler temperature. The force acting onto the MOT can be written ${\vec F}=-\alpha \vec{v}- \kappa_{x} \vec{x}- \kappa_{y} \vec{y}- \kappa_{z} \vec{z}$ at low velocities and close to the trap center.  The expected RMS cloud size $\sigma_{i}$ of MOT is then given by 
\begin{equation}
\frac{1}{2}\kappa_{i} \sigma_{i}^2 = \frac{1}{2} k_{B} T,
\label{Eq:MOTsize}
\end{equation}
with the three-dimensional equilibrium temperature $T$ calculated from Eq.~\ref{Eq:Tdoppler} and index $i$ referring to the different coordinate axis, $i=\{x,y,z\}$. The one-dimensional spring constants $\kappa_{i}$ are given by
\begin{gather}
\kappa_{i}=-\alpha_{\rm 2level} \frac{\mu B'_{i}} {\hbar k}
\end{gather}
where $B'_{i}$ is the magnetic field gradient along direction $i$ and $\mu=3 \mu_{B}/2$ is an average magnetic moment for the multi-level atom on the the $2^3 S_{1} \rightarrow 2^3P_{2}$ transition (Land\'e factors are $g_{J}(2^3S_{1})=2$ and $g_{J}(2^3P_{2})=3/2$). From this model we calculate the expected MOT sizes from the known experimental parameters. %

\begin{figure}[h!]
\begin{center}
\includegraphics[width=\columnwidth]{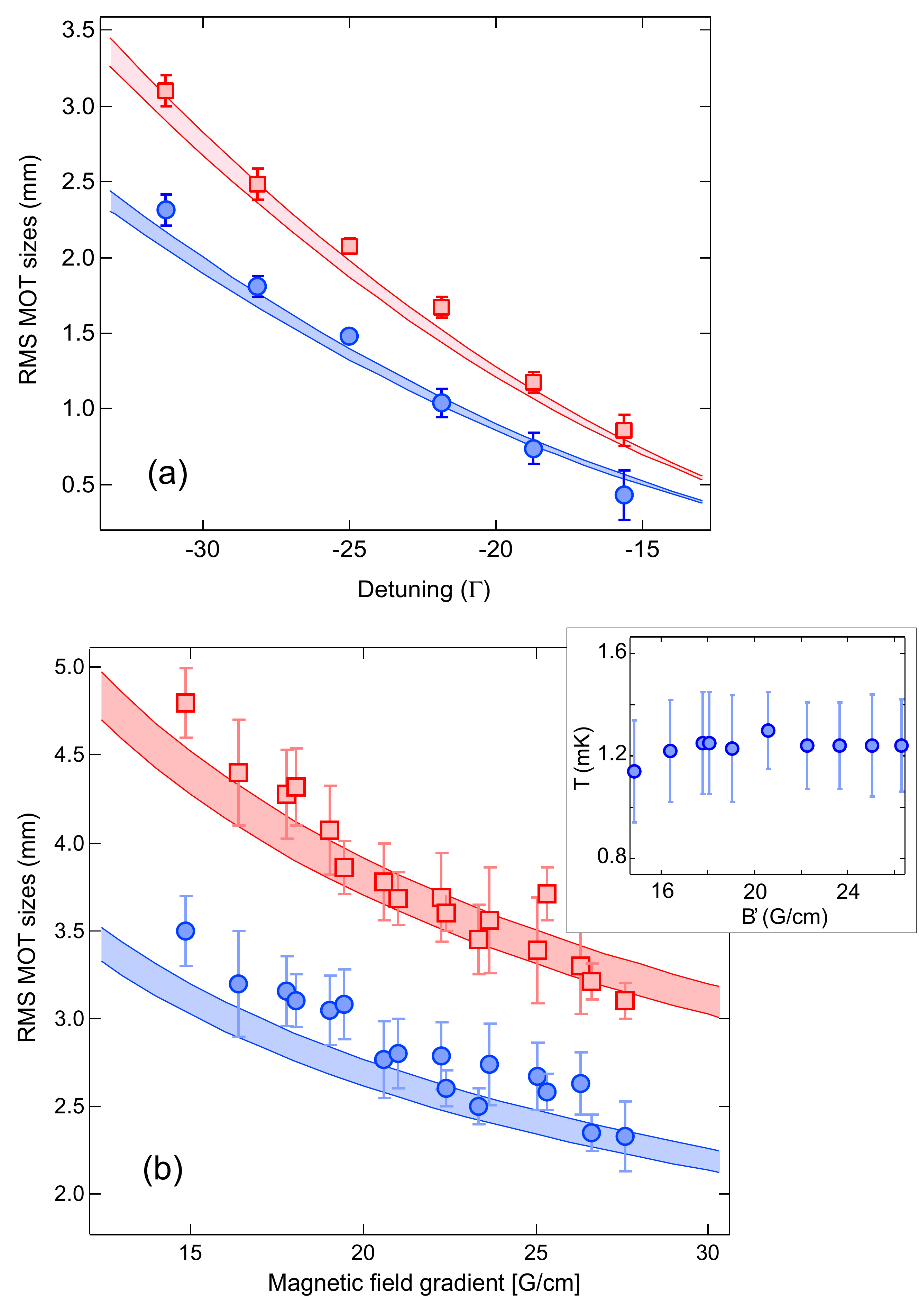}%
\caption{RMS sizes of the MOT clouds $\sigma_{x}$, $\sigma_{y}$ (blue, red) as a function of the laser detuning {\textbf (a)} and as a function of the magnetic field gradient {\textbf (b)}. The lines result from numerical calculation (no adjustable parameter) of the expected sizes in the regime of Doppler cooling (see text), accounting for a 5$\%$ error on the calibration of the light intensity.  Inset in {\textbf (b)} : temperature of the MOT as a function of the magnetic field gradient $B'$.}%
\label{SizeMOT}%
\end{center}
\end{figure}

We have measured the RMS MOT sizes as a function of both the axial magnetic-field gradient $B_{x}'$ and the detuning $\delta$ of the MOT beams. We plot in Fig.~\ref{SizeMOT} the experimental measurements along with the theoretical predictions in the Doppler cooling regime. The behavior of the MOT sizes with both the magnetic field gradient and the laser detuning are in good agreement with the prediction from Eq.~\ref{Eq:MOTsize} with no adjustable parameters, further demonstrating the validity of the Doppler model. We emphasize that the scaling of the MOT size with the detuning is different in the sub-Doppler regime where the temperature scales inversely proportional to $|\delta|$ \cite{Townsend1995}. As expected the MOT temperature is independent of $B'$, as shown in the inset of Fig.~\ref{SizeMOT}{\color{blue}a}. Previous studies of MOT sizes were either conducted in sub-Doppler regime of cooling \cite{Wallace1994,Kohns1993,Hope1993} or with alkaline-earth-metal atoms where high diffusion coefficients and multiple scattering prevent observations similar to ours \cite{Xu2002,Loo2004}. 

\subsection{Stability of Doppler molasses and drift velocity}\label{ssec:driftVelocity}

One experimental observation that puzzled researchers in the very first experiments with laser cooling was the long lifetime of the optical molasses \cite{Chu1985,Lett1989}. These measurements were associated with an unexpected robustness of the laser-cooling against an intensity imbalance between counter-propagating laser beams. The additional decay rate induced by this imbalance was measured to be 10 to 20 times lower than that predicted for Doppler molasses. This discrepancy was soon after explained by the presence of sub-Doppler cooling mechanisms \cite{Dalibard1989,Lett1989}. 

\begin{figure}
\begin{center}
\includegraphics[width=0.8\columnwidth]{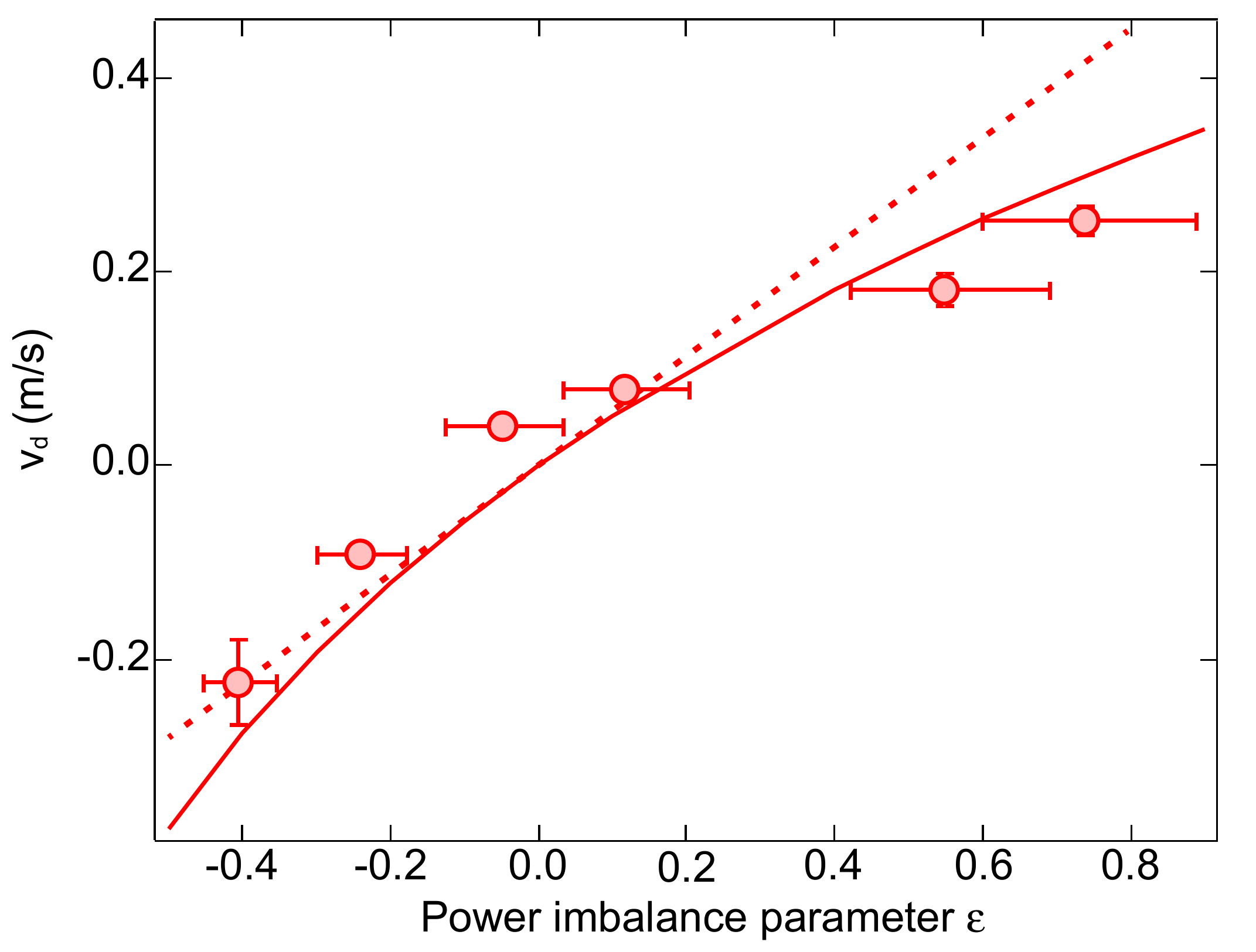}%
\caption{Drift velocity $v_{d}$ of the molasses cloud as a function of the power imbalance between two-counter propagating laser beams. For this measurement, we use $\delta=-\Gamma$ and $I/I_{sat}\simeq 1/40$.  The dashed line is the prediction of \cite{Lett1989} which is first order in $\epsilon$.  The solid line is the full prediction of Doppler theory, valid in the low-intensity limit.}%
\label{Fig:vdrift}%
\end{center}
\end{figure}

We have investigated the stability of the Helium molasses against intensity imbalance between counter-propagating laser beams.  Following \cite{Lett1989}, we define the imbalance parameter $\epsilon$ by $P_{1}=(1+\epsilon) P_{2}$, where $P_{1}$ ($P_{2}$) is the power in the forward-propagating (counter-propagating) beam along one given axis. We measure the drift velocity $v_{d}$ of the molasses cloud along this axis in the following way.  We repeat experimental cycle with varying molasses time $t_{mol}$, taking fluorescence pictures after a fixed and short time-of-flight $t_{TOF}=0.1$ ms. Recording the time evolution of the center of the molasses cloud, we extract a drift velocity $v_{d}$. In Fig.~\ref{Fig:vdrift} we plot the drift velocity as a function of the power imbalance between the two laser beams.

Following the derivation of the Doppler force valid in the limit of low intensity $s_0\ll 1$ \cite{Lett1989}, we derive the net force in the case of an unbalanced beam intensity.  In this case, the scattering forces from counter-propagating beams balance at a non-zero velocity $v_{d}$.  Although we obtain an analytical formula for $v_{d}$ under those circumstances, we express the result in orders of the imbalance $\epsilon$ to elucidate the physical content, 
\begin{equation}
v_{d}=\frac{\epsilon \Gamma}{8 k} \frac{1+ 4 \delta^2 / \Gamma^2}{2 | \delta| /\Gamma}-\frac{\delta\epsilon^2}{8k}\frac{1+4\delta^2/\Gamma^2}{4\delta^2/\Gamma^2}+\dots
\label{Eq:vDrift}
\end{equation}
In particular, the first order term of Eq.~\ref{Eq:vDrift} is found to be consistent with that reported in \cite{Lett1989} and results from writing the force $F=- \alpha_{2level} v$ with $\alpha_{2level}$ being expressed from Eq.~ \ref{Eq:alpha}. 

We plot in Fig.~\ref{Fig:vdrift} the results of the predictions for the intensity and detuning used in the experiment. The good agreement of the measurements with the Doppler prediction without adjustable parameters is an additional confirmation that the red optical molasses we manipulate are indeed in the regime of Doppler cooling, and that sub-Doppler cooling mechanisms are ineffective on the $2^3S_{1}\rightarrow 2^3P_{2}$ transition of $^4$He*.  Lastly, we note that similar behavior is observed for an imposed bias magnetic field.

\section{Conclusion}

In this paper we have reported on the first experimental observation of three-dimensional laser cooling at the Doppler limit with Helium-4 atoms. We have found quantitative agreement with the Doppler theory of (i) the temperature dependence as a function of the laser detuning, (ii) the MOT sizes and (iii) the velocity drift in optical molasses with unbalanced power between counter-propagating light beams. In particular we find the celebrated minimum Doppler temperature of $T_D=\hbar\Gamma/2k_B$ at a laser detuning of $\delta=-\Gamma/2$, as predicted in the simple picture of Doppler laser cooling.  We have shown that this behavior arises due to the special properties of metastable Helium atoms

The reported absence of sub-Doppler mechanism is related to the limited capture range allowed by the relatively low intensities required in red-detuned molasses. On the contrary, in blue-detuned molasses there is no such restriction, and high intensities can be used  to extend the capture range of sub-Doppler cooling \cite{Aspect1986, Grier2013}. This paves the way for the investigation of possible sub-Doppler cooling on the blue-detuned side of the $2^3S_{1} \rightarrow 2^3P_{1}$ transition with Helium-4.

\begin{acknowledgements}

We acknowledge fruitful discussions with D. Boiron, T. Bourdel, C. Cohen-Tanoudji and L. Sanchez-Palencia. We thank  A. Guilbaud, F. Moron, F. Nogrette and A. Villing, along with all members of the Atom Optic group at LCF, for technical help during the building of the experimental apparatus. We acknowledge financial support from the R\'egion Ile-de-France (DIM Daisy), the RTRA Triangle de la Physique (Junior Chair to D. C.), the European Research Council (senior grant Quantatop) and the Institut Francilien de Recherche sur les Atomes Froids (IFRAF). LCF is a member of IFRAF.

\end{acknowledgements}

\end{document}